\documentclass[aps,showpacs,showkeys,superscriptaddress]{revtex4}

\usepackage{graphicx}

\begin{document}
\title{Analogies in electronic properties of graphene wormhole and perturbed nanocylinder}

\author{J. Smotlacha}\email{smota@centrum.cz}
\affiliation{Bogoliubov Laboratory of Theoretical Physics, Joint
Institute for Nuclear Research, 141980 Dubna, Moscow region, Russia}
\affiliation{Faculty of Nuclear Sciences and Physical Engineering, Czech Technical University, Brehova 7, 110 00 Prague,
Czech Republic}

\author{R. Pincak}\email{pincak@saske.sk}
\affiliation{Bogoliubov Laboratory of Theoretical Physics, Joint
Institute for Nuclear Research, 141980 Dubna, Moscow region, Russia}
\affiliation{Institute of Experimental Physics, Slovak Academy of Sciences,
Watsonova 47,043 53 Kosice, Slovak Republic}

\date{\today}

\pacs{81.05.ue; 73.22.-f; 72.80.Vp}

\keywords{graphene wormhole, perturbed nanocylinder, Green function,
Fermi level, graphene blackhole}

\def\wu{\widetilde{u}}
\def\wv{\widetilde{v}}

\begin{abstract}
The electronic properties of the wormhole and the perturbed
nanocylinder were investigated using two different methods: the
continuum gauge field-theory model that deals with the continuum
approximation of the surface and the Haydock recursion method that
transforms the surface into a simplier structure and deals with the
nearest-neighbor interactions. Furthermore, the changes of the
electronic properties were investigated for the case of enclosing
the appropriate structure, and possible substitutes for the encloser
were derived. Finally, the character of the electron flux through the
perturbed wormhole was predicted from the model based on the
multiwalled nanotubes. The effect of the graphene blackhole is introduced.
\end{abstract}

\maketitle

\section{Introduction}The carbon nanostructures play a key role in constructing nanoscale devices like quantum wires, nonlinear
electronic elements, transistors, molecular memory devices or electron field emitters. Their molecules are variously-shaped geometrical forms whose surface is composed of disclinated hexagonal carbon lattice. The wormhole \cite{herrero} is created when two graphene sheets are connected through a small nanotube (so-called wormbridge) and through the singularities which emerge by adding 6 heptagonal defects to the connecting parts of the sheets with the wormbridge.

To characterize the electronic properties, the local density of states ($LDoS$) is investigated. For its calculation, the continuum gauge field-theory can be used in which the knowledge of the solution of the Dirac equation for the conduction electron is necessary \cite{mele}. It is represented by the wave-function, and to find it, we have to know the geometry of the corresponding surface. The Haydock recursion method \cite{haydock,tamura} transforms the surface into a chain of sites, each of them representing the equivalent sites in the original structure. The $LDoS$ is then acquired from the Green function which is calculated from an iterative formula \cite{tamura}.

The electronic flux can be influenced by a mechanical deformation of the surface by creating the surface-geometry induced attractive potential. In \cite{joglekar} is described, how to achieve this effect by a massive quantum particle present on a two-dimensional surface. This surface can be presented by a monolayer or a bilayer of graphene. The resulted potential suppresses the local Fermi energy. For this model, we can derive the relativistic dynamics and calculate the energy bands \cite{atanasov}. For the case of the deformation and the subsequent strain, this model is described in \cite{katenoid}. There is presented the geometry of the catenoid which connects the too sheets of the graphene. In this context of the strain induced potential, we can speak about so called "straintronics". Other possible geometries are the multiwalled nanotubes or fullerenes \cite{pds, pds2} or the deformed wormhole.

In this paper, we calculate the $LDoS$ of the wormhole using the mentioned methods, and we compare the results with the case of a perturbed nanocylinder including 2 heptagons at the opposite sides of the surface. It is organized as follows: the second section describes the metrics of the investigated nanostructures. In the third and the fourth section, the $LDoS$ of the wormhole and the perturbed nanocylinder is compared using the continuum gauge field-theory and the Haydock recursion method. Then, the term "perturbed wormhole" is introduced. In the fifth section, we investigate how to enclose the perturbed nanocylinder, and we look into the changes in the electronic structure. Next, the electron flux will be investigated using the model coming from the case of the multiwalled nanotubes.\\

\section{Wormhole and perturbed nanocylinder}
\begin{figure}
{\includegraphics{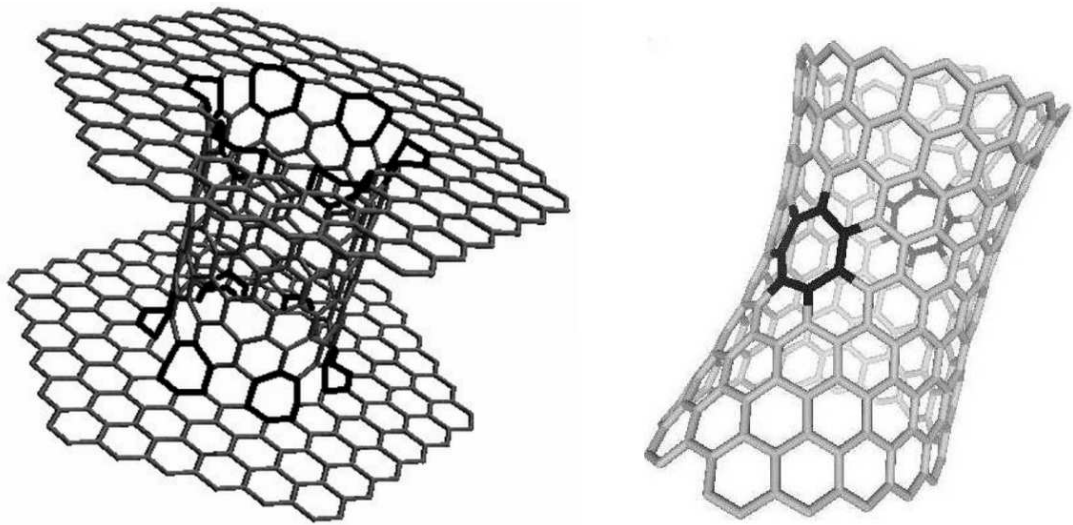}}\caption{Different surfaces derived from the cylindrical structure: wormhole (left), perturbed nanocylinder (right).}\label{fgmod}
\end{figure}

\begin{figure}
{\includegraphics{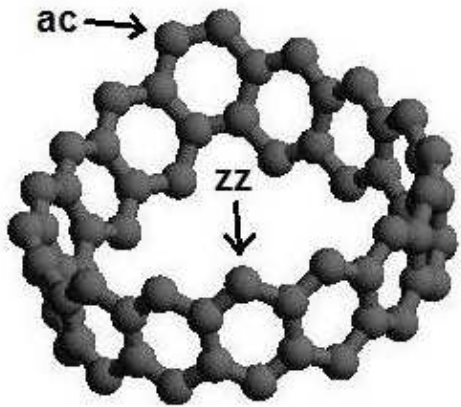}}\caption{Character of the edge corresponding to the perturbed nanocylinder: we cannot strictly say if the nanostructure is $ac$ or $zz$, it depends on the concrete position on the edge.}\label{fgedge}
\end{figure}

The surface of the investigated structure is depicted in fig. \ref{fgmod}. Contrary to the case of the wormhole, the perturbed nanocylinder contains only 2 heptagonal defects. It is derived from the defect-free nanocylinder which can be, similarly to the nanotubes, classified as armchair ($ac$), zig-zag ($zz$) and achiral. These 3 forms can be distinguished with the help of the chiral vector $(n,m)$ \cite{10}. All three forms differ by the shape of the edge.  But it is evident from fig. \ref{fgedge} that in the case of the perturbation, the shape of the edge changes along the circumference and that is why we cannot do such a classification for the case of the perturbation. We can only say which one of the 3 forms resembles a concrete site.

To investigate the physical properties of an arbitrary nanostructure, the knowledge of its metric is necessary. First, the radius vector has the form
\begin{equation}\label{8}\overrightarrow{\textbf{R}}(\xi,\varphi)=(x(\xi,\varphi),y(\xi,\varphi),
z(\xi,\varphi)),\end{equation} where $\xi$ and $\varphi$ are the coordinates with the help of which we parametrize the 2-dimensional surface embedded into 3 dimensions. Then, the metric is characterized by the metric tensor ${\bf g_{\mu\nu}},\hspace{2mm}\mu,\nu\in\{\xi,\varphi\}$, defined as ${\bf g_{\mu\nu}}=\partial_{\mu}\overrightarrow{\textbf{R}}\partial_{\nu}\overrightarrow{\textbf{R}}.$ The investigated cases are rotationally symmetric, so the non-diagonal components of the metric are $g_{\xi\varphi}=g_{\varphi\xi}=0.$

The radius vector of the perturbed nanocylinder has the form
\begin{equation}\label{sour}\overrightarrow{\textbf{R}}(z,\varphi)=\left(a\sqrt{1+\triangle z^2}\cos\varphi,
a\sqrt{1+\triangle z^2}\sin\varphi,z\right).
\end{equation}
Because of the structure of the cylinder, we use here the coordinate $z$ instead of $\xi$. The meaning of $a$ is the radius in the middle of the structure and $\triangle$ is a positive real parameter. For $\triangle\ll 1$, the components of the metric tensor will be
\begin{equation}\label{metric_pert}g_{zz}=1+1/(1+\triangle z^2)\sim 1+a^2\triangle^2 z^2,\hspace{2mm}
g_{\varphi\varphi}=a^2(1+\triangle z^2).\end{equation}

The wormhole geometry can be described by the polar-like coordinates denoted as $r_-,\varphi_-$ or $r_+,\varphi_+$, respectively, where $0<r_-,r_+<+\infty$. We choose the convention
\begin{equation}\label{inverse}r_-=a^2/r_+\end{equation}
where $a$ is the radius of the wormhole (it coincides with the radius of the nanocylinder), $r_-\geq a$ for the lower sheet and $r_+\geq a$ for the upper sheet, respectively.
Then, the corresponding metric tensor is \cite{herrero}
\begin{equation}\label{MT}{\bf g_{\mu\nu}}=\Lambda^2(r_{\pm})\left(\begin{array}{cc}1 & 0\\0 & r_{\pm}^2\end{array}\right),\end{equation}
where $\Lambda(r_{\pm})=\left(a/r_{\pm}\right)^2\theta (a-r_{\pm})+\theta (r_{\pm}-a),$
$\theta$ being the Heaviside step function. Because of (\ref{inverse}), the choice of the coordinates may seem to slant the real geometry: the meaning of $r_-,r_+$, respectively, on the opposite sheets, has nothing to do with the distance from the wormhole. But, by computing the Euler characteristics for the continuous surface, we get $\chi=\int{\rm d}^2x\sqrt{{\rm det}\,g}\mathcal{R}=-2,$ where $\mathcal{R}$ is the Ricci curvature. The acquired value is the same as the Euler characteristics of the corresponding carbon lattice. Next, we include an additional assumption that $a$, the radius of the wormhole bridge, is much larger than its length. This is the minimal model which describes the geometry of the wormhole \cite{herrero}.

\section{Continuum gauge field-theory}In this section, we determine the $LDoS$ from the solution of the Dirac equation in (2+1) dimensions. It has
the form
\begin{equation}\label{DirC}{\rm i}{\bf\sigma^{\alpha}e_{\alpha}^{\mu}}[\partial_{\mu}+\Omega_{\mu}-{\rm i}a_{\mu}-{\rm i}a_{\mu}^W]\psi=
E\psi.\end{equation}\\
The metric will be incorporated using the zweibeins $e_{\alpha}$ and the spin connection $\Omega_{\mu}$ \cite{1}. In the gauge field $a_{\mu}$, the influence of the present defects is included. For the case of the perturbed nanocylinder, if we denote their number by $N$, then $a_{\varphi}=N/4,\hspace{3mm}a_{\xi}=0.$ Here, we put $N=2$.

For the case of the wormhole, we have \cite{herrero} $a_{\varphi}=\Phi/(2\pi),\hspace{3mm}a_{\xi}=0,$
where $\Phi=-3\pi$ if the difference $n-m$ of the components in the chiral vector of the wormhole bridge is a multiple of $3$ and $\Phi=-\pi$ if the mentioned difference is not a multiple of $3$.

The gauge field $a_{\mu}^W$ is used only in the case of the perturbed nanocylinder. It is connected with the chiral vector $(n,m)$ of the defect-free structure from which the perturbed structure is derived, and the values of its components are $a_{\varphi}^W=-(2m+n)/3,\hspace{3mm}a_{\xi}^W=0.$ The meaning of all the other constituents in (\ref{DirC}) is described in \cite{1}.

The wave-function $\psi$ which solves (\ref{DirC}) has the form
\begin{equation}\label{wavef}\left(\begin{array}{c}\psi_A \\ \psi_B\end{array}\right)
=1/\sqrt[4]{g_{\varphi\varphi}}\left(\begin{array}{c}u(E,\xi){\rm e}^{{\rm i}\varphi
j}\\ v(E,\xi){\rm e}^{{\rm i}\varphi(j+1)}\\\end{array}\right),\hspace{2mm} j=0,\pm
1,...\end{equation}
where each of the components $\psi_A, \psi_B$ corresponds to one of two different sublattices $A, B$ of the hexagonal plane lattice \cite{wallace}. The introduced factorization of the solution will be substituted into (\ref{DirC}). Then we obtain
\begin{equation}\partial_{\xi}u/\sqrt{g_{\xi\xi}}-\widetilde{j}/
\sqrt{g_{\varphi\varphi}}\cdot u
=Ev,-\partial_{\xi}v/\sqrt{g_{\xi\xi}}-\widetilde{j}/\sqrt{g_{\varphi\varphi}}\cdot v=Eu,\end{equation}
where
$\widetilde{j}=j+1/2-a_{\varphi}-a_{\varphi}^W.$

For the given $\xi_0$, the $LDoS$ is defined as $LDoS(E)=|u(E,\xi_0)|^2+|v(E,\xi_0)|^2.$ In our calculations, the chiral vector of the perturbed nanocylinder will be $(12,0)$. In (\ref{wavef}), we choose the value $j=0$ for both the perturbed nanocylinder and the wormhole.

In the case of the wormhole, we get the solution of (\ref{DirC})
$u(r,E)=C_1(E)J_{\alpha}(E r)+C_2(E)Y_{\alpha}(E r),\hspace{3mm}v(r,E)=C_3(E)J_{\beta}(E r)+C_4(E)Y_{\beta}(E r),$
where
$\alpha=1/2\left|\Phi/\pi+1-2j\right|,\hspace{3mm}\beta=1/2\left|\Phi/\pi-1-2j\right|$
and $C_1(E),C_2(E),C_3(E),C_4(E)$ are such that the normalization is satisfied and it works for the initial values. $J_{\alpha}(x), J_{\beta}(x)$ and $Y_{\alpha}(x), Y_{\beta}(x),$ resp., are the Bessel functions of the first and the second kind, respectively.

Similarly, in the case of the perturbed nanocylinder, the solution is
$u(z,E)=C_1(E)D_{\nu_1}(\xi(z))+C_2(E)D_{\nu_2}({\rm i}\xi(z)),\hspace{3mm}
v(z,E)=C_1(E)/E\left(\partial_zD_{\nu_1}(\xi(z))-\widetilde{j}D_{\nu_1}(\xi(z))
(1-0.5\triangle^2z^2)/a\right)+C_2(E)/E\left(\partial_zD_{\nu_2}({\rm i}\xi(z))-\widetilde{j}D_{\nu_2}({\rm i}\xi(z))
(1-0.5\triangle^2z^2)/a\right),$ where $D_{\nu}(\xi)$ is the parabolic cylinder function \cite{12}, and the constants $\nu_1, \nu_2$ and the function $\xi(z)$ can be calculated from the input parameters. Again, $C_1(E),C_2(E)$ satisfy the normalization.

\begin{figure}
{\includegraphics{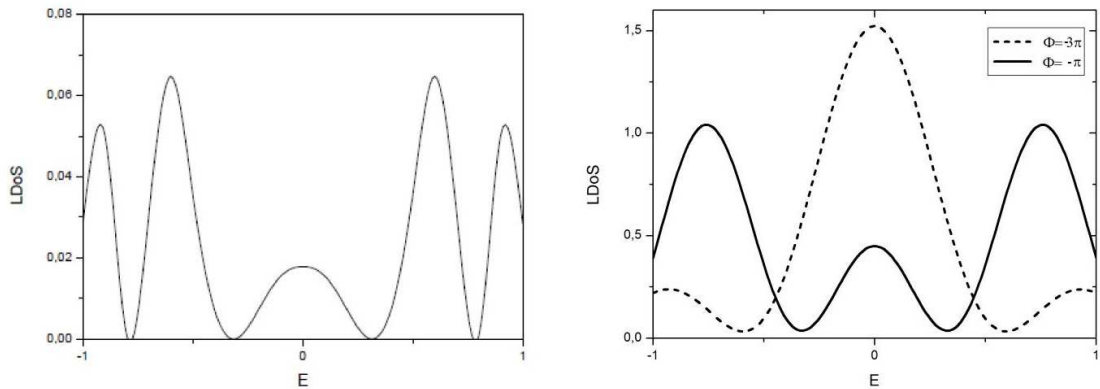}}\caption{$LDoS$
of the perturbed cylinder with $\triangle=0.05$ as a function of $E\in (-1,1)$ on the edge site (left) and of the wormhole (right).}\label{fg4}
\end{figure}

In fig. \ref{fg4}, the local density of states of the wormhole and of the perturbed nanocylinder on the edge site is compared. We see that in the case $\Phi=-\pi$ (the difference $n-m$ of the coordinates in the chiral vector is not a multiple of $3$), the results are very similar for both the cases.\\

\section{Haydock recursion method}In this section, we give a short description of the Haydock recursion method. As stated in the Section I, this method transforms the surface into a chain of sites each of them represents the equivalent sites in the original structure. The results acquired by this method are more precise than the results acquired using the continuum gauge field-theory \cite{green}.

The sites are represented by the state vectors $|n\rangle, n=1,\ldots,n_{max}$. From
the action of the Hamiltonian corresponding to the nearest-neighbor interaction follows \cite{tamura}
\begin{equation}\label{1}H|n\rangle=a_n|n\rangle+b_{n-1}|n-1\rangle+|n+1\rangle,\end{equation}
where $a_n, b_n, n=1,\ldots,n_{max}$ are real coefficients. Then, the $LDoS$ is defined as
\begin{equation}\label{LDOS}LDoS(E)=\lim\limits_{\delta\rightarrow +0}{\rm Im}G_{00}(E-{\rm i}\delta)/\pi,\end{equation}
where $G_{00}(E)$ is the Green function which will be calculated recursively using the procedure described in \cite{tamura}.

\begin{figure}
{\includegraphics{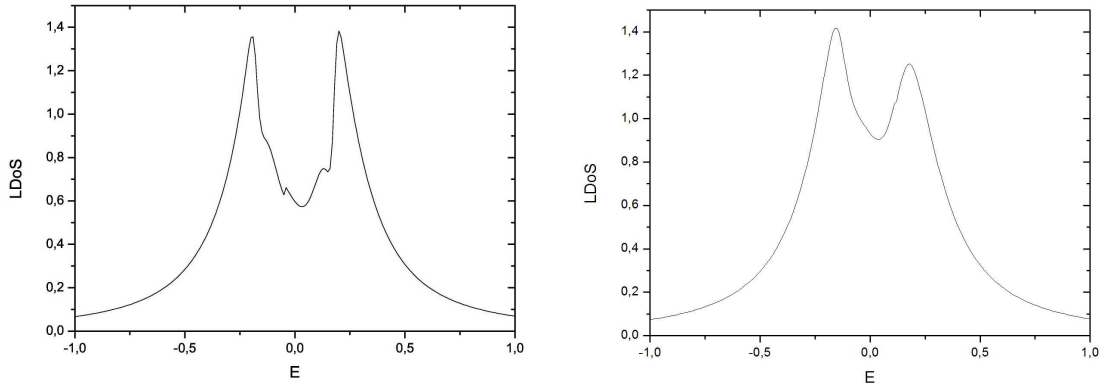}}\caption{$LDoS$ calculated using the Haydock recursion method for the perturbed nanocylinder (left) and the wormhole (right); here, $\delta=0.2$.}\label{fgN}
\end{figure}

In fig. \ref{fgN}, similarly as in the previous chapter, we see a comparison of the character of the $LDoS$ calculated by using this method for both the wormhole and the perturbed nanocylinder. It follows from the plot that, analogously to the case of using the continuum gauge field-theory for the calculation of the $LDoS$ of the perturbed nanocylinder and of the wormhole with $\Phi=-\pi$, the results are similar for both the cases.\\

\section{Perturbed wormhole}So in the following we will speak about the perturbed wormhole instead of the perturbed nanocylinder. By the perturbed wormhole we will understand the structure which will be similar to the wormhole, but the curvature will not be established by $12$ heptagonal defects, as in the case of the wormhole, but it will be mediated by only $2$ heptagonal defects which will be placed in the same way as in the case of the perturbed nanocylinder. The resulting nanostructure arises by adding the graphene structure to the edges of the perturbed nanocylinder. This creates a continuous prolongation whose form could be similar to the Beltrami pseudosphere \cite{beltrami}, but the mechanical deformation causes the adaptation to the definitive form. One of possible parametrizations (logarithmic) of the resulting surface can be found \textit{e.g.} in \cite{katenoid}. In this paper, the deformation will be described by the parameter $\triangle$ which appears in (\ref{sour}). We will suppose that, on the contrary to the calculations made in Section III, the chiral vector will have different components than $(12,0)$. The reason is that for this case, the difference $n-m$ is a multiple of $3$ and as follows from fig. \ref{fg4}, for this case the value of $\Phi$ in the wormhole is $-3\pi$. The corresponding plots of $LDoS$ for both structures would not be then similar. On the other hand, the results for the perturbed nanocylinder are not changed in the case of small changes of the chiral vector.\\

\section{Enclosure of the deformed structure}The electronic properties can be changed if we enclose the investigated structure by a nanostructured surface which contains some pentagonal defects. We will demonstrate this effect in the case of the perturbed wormhole.

First, we find how many pentagonal defects $N(\triangle)$ must be present in the enclosing structure and investigate the geometry of this structure. After doing this and the calculation of $N(\triangle)$, we find the value of $\triangle$ which is needed to use some concrete forms of the fullerene molecules as the encloser. Then, we investigate how the energy of the "Fermi levels" of the infinitely small nanotubes, from which the perturbed wormhole is composed (see the Subsection C for the detailed explanation), depends on the distance from the heptagonal defect.\\

\begin{figure}
{\includegraphics{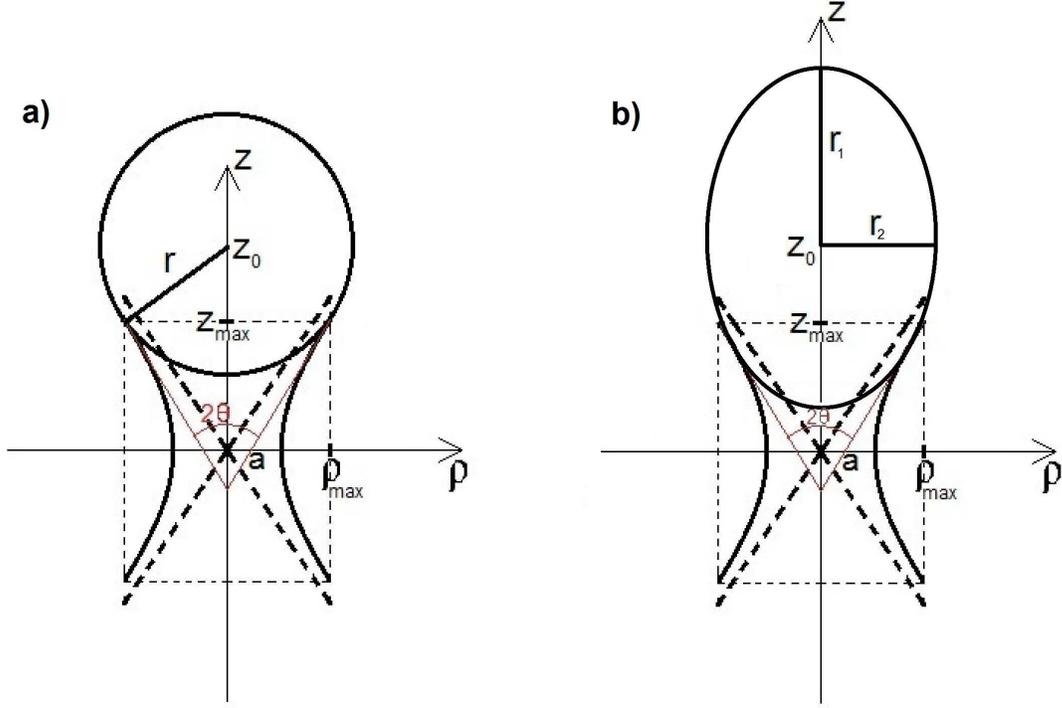}}\caption{Perturbed wormhole enclosed by the spherical (a) or rotationally elliptical surface (b).}\label{fgD}
\end{figure}

\subsection{Geometry and included defects}The investigated structure is depicted in fig. \ref{fgD}a. (Fig. \ref{fgD}b shows that this structure can be more complicated but this case will not be investigated here.) In this figure, we see the perturbed wormhole which is enclosed by a spherical surface of the radius $r$ which encloses the structure. The coordinates of the perturbed wormhole surface will be denoted by ($\rho_w, z_w$) and the coordinates of the surface of the encloser will be denoted by ($\rho_s, z_s$). Both the surfaces are connected in the position given by the coordinates ($\rho_{max},z_{max}$).

The deformation of the wormhole is described by the parameter $\triangle$ and from (\ref{sour}) the relation follows between the coordinates $z_w$ and $\rho_w$:
\begin{equation}\label{cleq1}\rho_w(z_w)=a\sqrt{1+\triangle z_w^2},\end{equation}
where $\rho_w^2=x_w^2+y_w^2$ and $a$ is the radius of the center of the perturbed wormhole bridge.

The sphere is described by
\begin{equation}\label{cleq2}z_s-z_0=\pm\sqrt{r^2-\rho_s^2},\end{equation}
where the sign $"\pm"$ corresponds to the top and to the bottom part of the sphere, respectively. The corresponding sign for the position ($\rho_{max}, z_{max}$) is $"-"$. The parameters $r, z_0$ can be calculated from the requirement of the connection of both surfaces in this position and of the continuity of the derivations: we have ${\rm d}z_w/{\rm d}\rho_w|_{\rho_{max}}={\rm d}z_s/{\rm d}\rho_s|_{\rho_{max}},$ from which follows after some modifications
\begin{equation}\label{cleq3}r=a\sqrt{1+\triangle z_{max}^2+a^2\triangle^2z_{max}^2}.\end{equation}
Now with the help of (\ref{cleq2}),(\ref{cleq3}) and (\ref{cleq1}) we can derive
\begin{equation}\label{37}z_0=z_{max}(1+\triangle a^2).\end{equation}

Now we can find $N(\triangle)$, the number of the defects which are needed to enclose the perturbed wormhole. As it is known from the Euler theorem, each enclosed structure, its defects are created by pentagons, contains exactly 12 defects. We denote by $N_d$ the number of the defects contained in the bottom part of the enclosing nanostructure. Then $N(\triangle)+N_d=12.$ Because the angle between the tangential lines is $2\theta$, it follows from \cite{10} that $\sin\theta=1-N_d/6=N(\triangle)/6-1.$
The value of $\theta$ is between $0$ and $\pi/2$, so we easily see that the values of $N(\triangle)$ are between $12$ and $6$. We derive now how $N(\triangle)$ depends on a concrete value of $\triangle$.

It follows from the sketch in fig. \ref{fgD} that ${\rm d}z_w/{\rm d}\rho_w|_{\rho_{max}}=\tan\left(\pi/2-\theta\right)=\cot\theta$
and after using some identities and substituting $\rho_{max}=a\sqrt{1+\triangle z_{max}^2}$ we get
\begin{equation}N(\triangle)=6\left(1+a\triangle z_{max}/\sqrt{1+\triangle z_{max}^2+a^2\triangle^2z_{max}^2}\right).\end{equation}

\begin{figure}
{\includegraphics{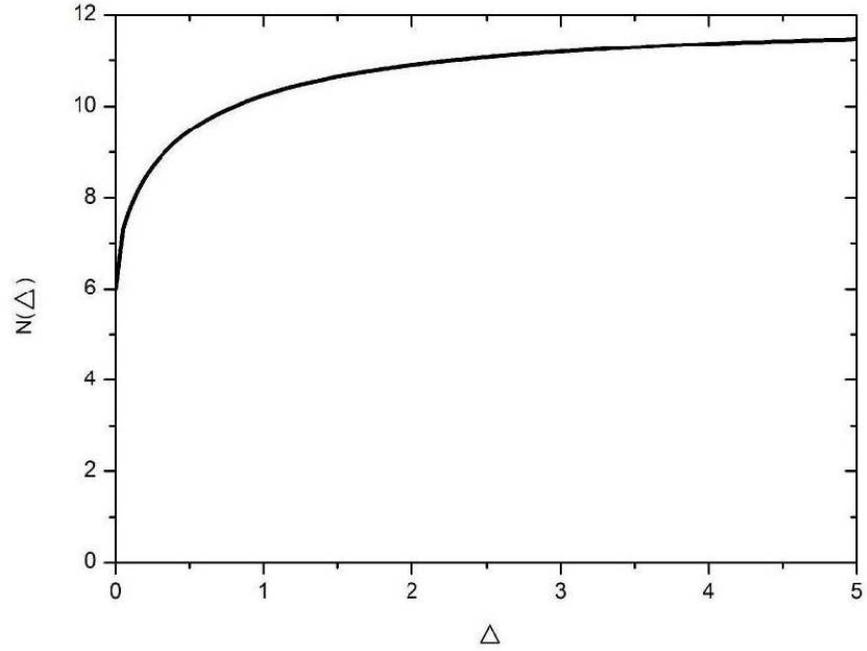}}\caption{Number of pentagonal defects which enclose the perturbed wormhole surface as a function of the parameter $\triangle$.}\label{fgND}
\end{figure}

\begin{figure}
{\includegraphics{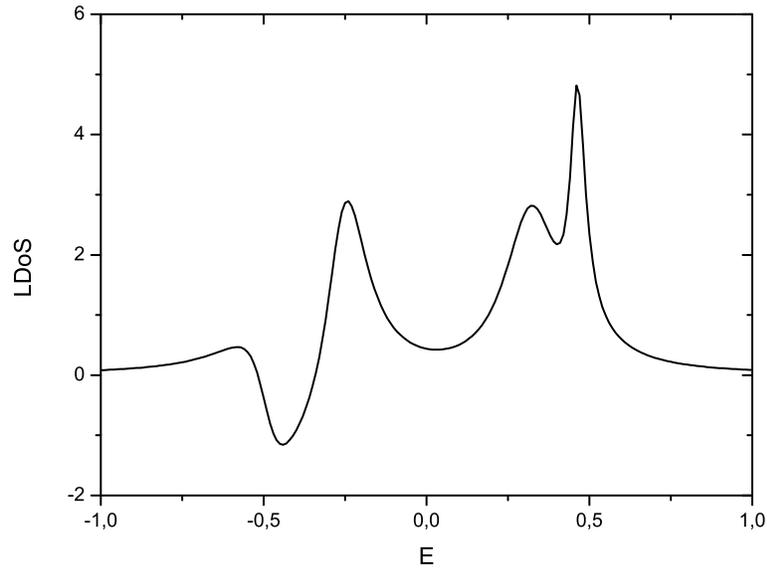}}\caption{$LDoS$ of the enclosed perturbed wormhole calculated using the Haydock recursion method; here, $\delta=0.2$.}\label{fguz}
\end{figure}

In fig. \ref{fgND}, we see how the number of needed defects depends on the parameter $\triangle$. To investigate the change of the electronic properties, we will use the Haydock recursion method described in the previous section. The investigation will be carried out for the sites placed in the connecting part of the perturbed wormhole and the enclosing nanostructure (coordinates $\rho_{max}, z_{max}$ in fig. \ref{fgD}). The result is in fig. \ref{fguz}.\\

\subsection{Possible forms of fullerene molecules in the encloser}Now we find which form of the fullerene molecule can enclose the given structure with the given value of $\triangle$. This form is characterized by the ratio $d/r$, where $d$ is the length of the bond between the carbon atoms. For the fullerene $C_{60}$ \cite{kroto}, the length of the circumference corresponds to $p_{60}=15$ bonds, so approximately
$2\pi r=p_{60}d_{60},$ where $d_{60}$ denotes the length of the corresponding bond. Simultaneously, we fix the number of the atoms on the connection part of the perturbed wormhole as $15$. So in the case $\triangle=0$, the structure will be enclosed by the fullerene $C_{60}$ and, for the arbitrary deformation, in the case of the spherical surface,
$2\pi\rho_{max}=2\pi a\sqrt{1+\triangle z_{max}^2}=p_{60}d.$
Simultaneously, (\ref{cleq3}) holds, so after the substitution of the dimensionless parameter $\widetilde{\triangle}=\triangle\cdot a^2$,
\begin{equation}\begin{array}{c}\label{sph2}d/r=
2\pi/p_{60}\sqrt{\left(1+\widetilde{\triangle}(z_{max}/a)^2\right)}\cdot\\\\
\cdot\sqrt{1/\left(1+\widetilde{\triangle}(z_{max}/a)^2+
\widetilde{\triangle}^2(z_{max}/a)^2\right)}.\end{array}
\end{equation}
Now, we find the relation between $d/r$ and $\widetilde{\triangle}$ for some concrete cases of using the fullerene molecules as the encloser. But there are too many parameters and that is why we fix the ratio $z_{max}/a$. In the following calculations, we put $z_{max}/a=2$. For the cases of other values of this ratio, we would have to do other calculations. In table \ref{TabS}, the values of $d/r$ and $\widetilde{\triangle}$ are introduced for some kinds of the fullerene molecules.\\

\begin{table}
\caption{The values of $d/r$ and $\widetilde{\triangle}$ for different kinds of spherical surfaces present in the encloser.}
\begin{tabular}{@{\extracolsep{0.2cm}}cc|rrrrr}
\hline
  &  & $C_{60}$ & $C_{80}$ & $C_{180}$ & $C_{240}$ &\\
\hline
& $d/r$\hspace{0.2cm} & $0.419$ & $0.363$ & $0.242$ & $0.209$ &\\

& $\widetilde{\triangle}$\hspace{0.2cm} & $0$ & $0.498$ & $2.221$ & $3.249$ &\\
\hline
\end{tabular}\label{TabS}
\end{table}

\subsection{"Fermi levels" of the perturbed wormhole}
\begin{figure}
{\includegraphics{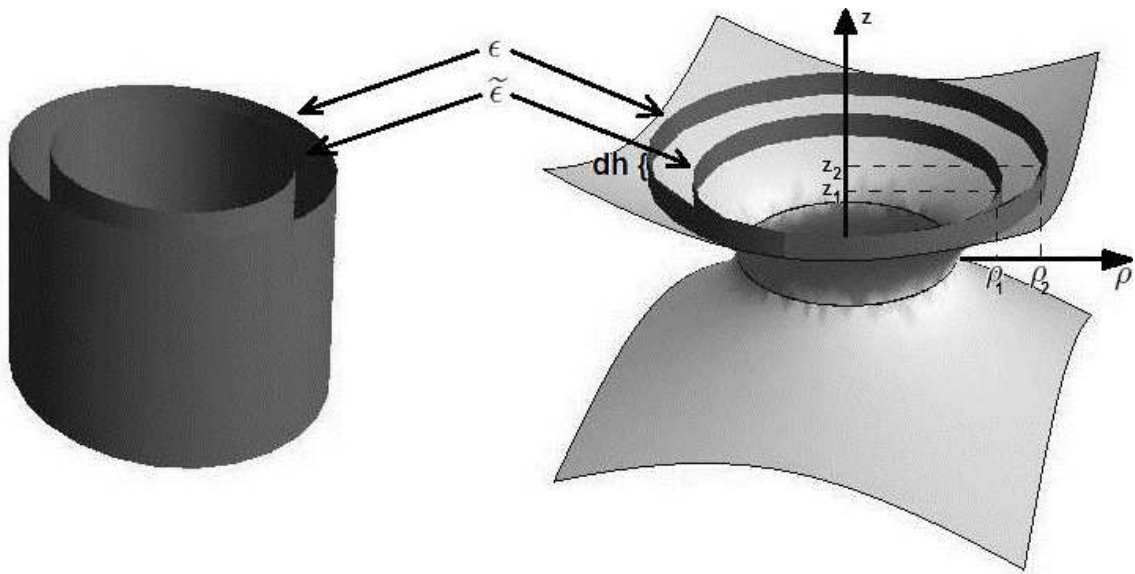}}\caption{Similarity of the structures of the multiwalled nanotubes and the perturbed wormhole: the thickness of the wormhole sheet has a very small value, we denote it by d$h$. Then, the perturbed wormhole can be understood as a composition of very low and thin nanotubes. In this way, we can calculate the "Fermi levels" at each "nanotube". The meaning of the particular symbols is explained in the text.}\label{Fermi}
\end{figure}

The concentric circles of which the sheets of the perturbed wormhole are composed can be understood as very low and thin nanotubes which are ordered very close to each other (see fig. \ref{Fermi}). This can be exploited for the investigation of the effect which was proven in \cite{pds, pds2} for the case of the multiwalled nanotubes (or the multiwalled fullerenes, respectively): the Fermi level of the electrons on the outer nanotubes (with the higher radius) is higher than the Fermi level of the electrons on the inner nanotubes (with the lower radius). Then, in a similar way we can formally calculate the difference of the "Fermi levels" as \cite{pds}
\begin{equation}\label{levdif}\begin{array}{c}\epsilon-\widetilde{\epsilon}=\pi^2(1+4\widetilde{\triangle})
\left((z_1/a)^2-(z_2/a)^2\right)\widetilde{\triangle}\cdot\\\\
\cdot 1/\left(36l_c^2(1+z_1^2\widetilde{\triangle}/a^2)
(1+z_2^2\widetilde{\triangle}/a^2)\right)\cdot\\\\
\cdot\left(2\langle s|H|s\rangle+\langle p|H|p\rangle\right)\end{array}\end{equation}
($\epsilon,\widetilde{\epsilon}$ correspond to the "Fermi level" of the outer and the inner circle, respectively, $\langle s|H|s\rangle, \langle p|H|p\rangle$ are the energies of the corresponding $s$ and $p$ orbitals \cite{pds}; $z_1, z_2$ are $z$ coordinates of the circles). Here, the expression "Fermi level" is written in the quotation marks, because the bonds between the particular circles are much stronger than in the case of the maltiwalled nanotubes and in fact, in the case of the precise calculations, they cannot be taken as the separated structures. What we present here, is a rough approximation.

We choose some fixed values of $z_1, z_2$ and we will compare the difference of the "Fermi levels" of the circles for different deformations, \textit{i.e.} for different values of $\widetilde{\triangle}$. We use the values from table \ref{TabS}. So we put $z_1=z_{max},\hspace{3mm}z_2=1.2z_{max},\hspace{3mm}l_c=15;$ then $z_1/a=2,\hspace{3mm}z_2/a=2.4$
and taking into account that \cite{pds} $\langle s|H|s\rangle=-12\,{\rm eV },\hspace{3mm}\langle p|H|p\rangle=-4\,{\rm eV },$
we get the difference of the "Fermi levels", as introduced in table \ref{Tab3}.\\

\begin{table}
\caption{The difference of the "Fermi levels" on the perturbed wormhole for the chosen values of $z_1, z_2$. The values of $\widetilde{\triangle}$ correspond to the values from table \ref{TabS}.}
\begin{tabular}{@{\extracolsep{5mm}}l|rrrr}
\hline
\hspace{0.5cm}$\widetilde{\triangle}$\hspace{0.5cm} & $0$ & $0.498$ & $2.221$ & $3.249\hspace{0.5cm}$\\

\hspace{0.5cm}$\epsilon-\widetilde{\epsilon}$\hspace{0.5cm} & $0$ & $0.023$ & $0.029$ & $0.030\hspace{0.5cm}$\\
\hline
\end{tabular}\label{Tab3}
\end{table}

\section{Conclusion}The comparison of the $LDoS$ of the wormhole and of the perturbed nanocylinder was performed using different methods. Both methods provided much different results, but in the case of the difference $n-m$ of the components of the chiral vector of the wormhole bridge not being a multiple of $3$, each of the methods confirmed similarity of both structures from the perspective of the electronic properties. In a different way, the equivalence of both structures was proven in \cite{dandol}. The value of the perturbation in the investigated structures was not very large and that is why we can compare our results with the calculations in some earlier works \cite{11,CN}. We can also make a comparison with the calculations performed for the capped nanotubes \cite{vanhove}.

As mentioned in Section II, the radius of the wormhole bridge is much larger than its length. Contrary to this, this assumption is not needed in the case of the introduced perturbed wormhole which theoretically can have a macroscopic size.

The similarity of the physical properties of the wormhole and of the perturbed nanocylinder can be exploited in many applications from the fields of nanoelectronics and nanooptics. On the other hand, the perturbed nanocylinder can be used as the substitute for studying astrophysical phenomena related to the gravitational effects connected with the electron quasiparticles.

If we enclose the given structure by a concrete number of pentagonal defects, we achieve a significant change of the electronic structure (see figs. \ref{fgN}, \ref{fguz}). Not only the investigated spherical surface (fig. \ref{fgD}a, table \ref{TabS}) is possible for the encloser. Other forms like \textit{e.g.} elliptical surface presented in fig. \ref{fgD}b can be investigated in a similar way.

The rise of the "Fermi levels" shows an important property of the related structures: the electron flux is directed from the far areas of the perturbed wormhole to the center. As a consequence, the electrical charge is accumulated in the center and in this way, we can speak about the so-called graphene blackhole. Detailed explanation of the related effects is given in \cite{atanasov}, where the effects accompanying the deformation of the graphene are described: the distance of the carbon atoms in the layer is changed. Next, the rotation of the $p_z$ orbitals occurs and the $\pi$ and $\sigma$ orbitals are rehybridizated . This procedure leads to the creation of the $p-n$ junctions similarly as in the case of transistor. By this way, the direction of the electron flux is influenced. The idea of the graphene blackhole in the case of the deformed wormhole is based on this effect which changes the Fermi level. It is rising in the far areas from the wormhole center and as a result the electron flux is directed from these areas to the middle where the electric charge is accumulated. The form of the nanotube in the middle plays a big role for this purpose. It can't be unperturbed because in such a case the effect of the blackhole would be disrupted. It can be ensured only in the case when the nanotubular neck is tapering in the direction to its center, because this ensures the decrease of the Fermi level \cite{pds, pds2}. The related effects appearing on the nanostructures are also described in \cite{beltrami}.

The effect of the graphene blackhole could eventually disappear in the presence of the external magnetic (electric) field which would cause the transfer of the charge from one of the wormhole sheets to another through the center. This serves as an important model for further investigations of the electron flux in the presence of the defects. In \cite{charge}, some investigations were carried out for the above mentioned wormhole with $12$ heptagonal defects. Possible investigations in the case of the next deformations could contribute to the applications in the cosmological models.

Unfortunately, we did not find any deeper conception which suggests a method of the production of the graphene wormholes. So, we think that it could be manufactured from the graphene bilayers whose properties are described for example in \cite{mucha} or \cite{jernigan}. We consider that the graphene monolayers could be mechanically pressed against each other so that their distance would be reduced below the value of the length of the atomic bonds in the graphene. Under these conditions, the interaction between the valence electrons of the carbon atoms from the opposite layers could achieve significant values, because it would exceed the interaction between the neighbors in the hexagonal carbon structure. Furthermore, as we considered above, the radius of the wormhole must be much longer than its length, so, the minimal distance between the monolayers is very important. The structure of the wormhole could then arise spontaneously.

In the paper were also found very similar analogies between the investigated structures. It is very important for the real applications in electronic nanodevice because the size of graphene wormhole is microscopic in contrary with the perturbed nanotube which can have macroscopic size.

\acknowledgments

ACKNOWLEDGEMENTS --- The work was supported by the Slovak Academy of
Sciences in the framework of CEX NANOFLUID, and by the Science and
Technology Assistance Agency under Contract No. APVV-0509-07 and
APVV-0171-10, VEGA Grant No. 2/0037/13 and Ministry of Education
Agency for Structural Funds of EU in frame of project 26220120021,
26220120033 and 26110230061. R. Pincak would like to thank the TH
division in CERN for hospitality.


\end{document}